\documentclass[12pt]{iopart}

\begin{document}

\title[Dalgarno-Lewis Method Revisited]{Dalgarno-Lewis Method Revisited}

\author{A B Balantekin and A Malkus}

\address{Department of Physics, University of Wisconsin, Madison, WI 53706 USA}
\eads{\mailto{baha@physics.wisc.edu}, \mailto{amalkus@wisc.edu}} 
\begin{abstract}
Proving the existence of an operator that connects non-perturbed states to perturbed states,  an alternative derivation of the Dalgarno-Lewis method is given.  To illustrate that the Dalgarno-Lewis method is an apt tool  for algebraic Hamiltonians, the method is applied to one class of such systems, namely  deep three-dimensional potentials with positive parity.  
\end{abstract}

\pacs{03.65.Fd,03.65.Ge}


\section{Introduction}

In perturbation theory the total Hamiltonian of the system under consideration is taken to be of the form 
\begin{equation}
\label{1}
\hat{H} = \hat{H}_0 + \lambda \hat{h},
\end{equation}
where it is assumed that the eigenstates and the eigenvalues of the Hamiltonian, $\hat{H}_0$, are 
known:  
\begin{equation}
\label{2}
\hat{H}_0 |n \rangle = \epsilon_n |n \rangle 
\end{equation}
and the parameter $\lambda$ is small. In this work we will assume that the eigenvalue spectrum of the operator $\hat{H}_0$ is non-degenerate. One is interested in finding the eigenvalue spectra of the operator $\hat{H}$:
\begin{equation}
\label{3}
\hat{H} | \Psi_n (\lambda )\rangle = E_n (\lambda)  |\Psi_n (\lambda) \rangle .
\end{equation}

The question we wish to address here is if there exists an operator $\hat{S}_n (\lambda) $ that connects each state in (\ref{2}) and (\ref{3}):
\begin{equation}
\label{4}
| \Psi_n (\lambda) \rangle = \hat{S}_n (\lambda)  |n \rangle 
\end{equation} 
and describe the connection between such operators and the Dalgarno-Lewis method \cite{DL}. 
Clearly one needs to impose 
\begin{equation}
\label{5}
\lim_{\lambda \rightarrow 0} \hat{S}_n (\lambda) = \hat{1}. 
\end{equation}
We assume that the operator $\hat{S}_n (\lambda)$ can be written in the form 
\begin{equation}
\label{6}
\hat{S}_n (\lambda) = \exp \left( \lambda \hat{F}_n + \lambda^2 \hat{G}_n + 
\lambda^3 \hat{\Gamma}_n\cdots \right),
\end{equation}
where $\hat{F}_n$, $\hat{G}_n$ and $\hat{\Gamma}_n, \cdots$ are operators to be determined. Inserting (\ref{4}) and (\ref{6}) into 
(\ref{3}), one finds that the condition
\begin{equation}
\label{7}
\left( [\hat{S}_n (\lambda)]^{-1} \hat{H} (\lambda)  \hat{S}_n (\lambda) \right) |n \rangle = E_n(\lambda) 
|n \rangle 
\end{equation}
needs to be satisfied. Expanding the eigenvalue, $E_n (\lambda)$, in powers of $\lambda$:
\begin{equation}
\label{8}
E_n (\lambda) = \epsilon_n + \lambda \epsilon_n^{(1)} + \lambda^2 \epsilon_n^{(2)} + \cdots, 
\end{equation}
one inserts into (\ref{7}). After equating powers of $\lambda$,  one gets the following conditions: 
\begin{eqnarray}
{\cal O}(\lambda):  \hat{h} - [\hat{F}_n,\hat{H}_0] =   \epsilon_n^{(1)} 
\label{9}  \\
{\cal O} (\lambda^2):  \frac{1}{2} [\hat{F}_n, [ \hat{F}_n, \hat{H}_0]] - [ \hat{G}_n, \hat{H}_0] 
- [ \hat{F}_n, \hat{h}] =   \epsilon_n^{(2)} \label{10} \\
{\cal O}(\lambda^3): - [\hat{G}_n, \hat{h}] - [\hat{\Gamma}_n, \hat{H}_0 ] + 
\frac{1}{2} [\hat{F}_n, [\hat{G}_n, \hat{H}_0]] + \frac{1}{2} [\hat{G}_n, [\hat{F}_n, \hat{H}_0]]  
\nonumber \\
+ \frac{1}{2} [\hat{F}_n, [\hat{F}_n, \hat{h}]] - \frac{1}{6} [ \hat{F}_n, [\hat{F}_n, [ \hat{F}_n, \hat{H}_0 ]]] 
= \epsilon_n^{(3)} \label{11} 
\end{eqnarray}
and so on. The equalities in (\ref{9}), (\ref{10}), and (\ref{11}) are understood to be acting on the state 
$|n \rangle$, e.g. (\ref{9}) is a short-hand notation for $ (\hat{h} - [\hat{F}_n,\hat{H}_0]) |n \rangle =   \epsilon_n^{(1)} |n \rangle$. This equation suggests that we can take the operators $\hat{F}_n$ to be the same for all $n$ and the part of that operator which is diagonal in the basis of the eigenfunctions of 
$\hat{H}_0$ drops out from (\ref{9}). Indeed multiplying (\ref{9}) by the state $\langle m|$ from left we find that
\begin{equation}
\label{12}
\langle m | \hat{h} | n \rangle = (\epsilon_n - \epsilon_m) \langle m | \hat{F} | n \rangle + 
\epsilon_n^{(1)} \delta_{nm} .
\end{equation}
For $n=m$ this equation yields the standard expression for the first order correction to the energy eigenvalues in non-degenerate perturbation theory. For $n\neq m$, (\ref{12}) gives non-diagonal elements of $\hat{F}$. Without any loss of generality we will take the diagonal elements of this operator 
to be zero and write
\begin{equation}
\label{13}
\hat{F} = \sum_{n \neq m} \left( \frac{\langle m | \hat{h} | n \rangle}{\epsilon_n - \epsilon_m} \right)
| m \rangle 
\langle n | . 
\end{equation}
We can now consider (\ref{9}) an operator relationship provided that the right-hand side is replaced by 
a diagonal operator in basis of the eigenstates of $\hat{H}_0$, the eigenvalues of which are 
$\epsilon_n^{(1)}$. Following similar arguments we can also consider (\ref{10}) and (\ref{11}) as operator relationships  provided that we replace the right-hand sides with appropriate diagonal operators. We will denote these operators $\epsilon^{(1)}, \epsilon^{(2)}, \epsilon^{(3)}, \cdots$. 
We find it  useful to spell out the diagonal and non-diagonal parts of the operator $\hat{h}$:
\begin{equation}
\label{14}
\hat{h} = \hat{h}_{\rm D} + \hat{h}_{\rm ND}.
\end{equation}
From (\ref{9}) we observe that $\hat{h}_{\rm D} = \epsilon^{(1)}$ and 
\begin{equation}
\label{15}
[\hat{F}, \hat{H}_0 ] = \hat{h}_{\rm ND}. 
\end{equation}
Using (\ref{15}), the operator equivalent of (\ref{10}) can be rewritten as
\begin{equation}
\label{16}
- [ \hat{G}, \hat{H}_0] - [ \hat{F}, \hat{h}_{\rm D}] - \frac{1}{2} [\hat{F}, \hat{h}_{\rm ND}] = 
\epsilon^{(2)}. 
\end{equation}
Inspection reveals that the first two commutators in (\ref{16}) yield non-diagonal operators, 
consequently one gets the standard second order result: 
\begin{equation}
\label{17} 
- \frac{1}{2}  \langle n| [\hat{F}, \hat{h}_{\rm ND}] | n \rangle = \epsilon_n^{(2)} .
\end{equation} 
Dalgarno and Lewis developed a method for calculating second-order corrections to the perturbation expansion, which essentially uses the same commutator \cite{DL}. 
The non-diagonal elements of the operator $\hat{G}$ can be calculated by taking the expectation value 
of (\ref{16}) between states $\langle m |$ and $|n \rangle$ ($m \neq n$): 
\begin{equation}
\label{18}
 \langle m | \hat{G} | n \rangle = - \frac{(\epsilon_n^{(1)} - \epsilon_m^{(1)})}{\epsilon_n -\epsilon_m} 
  \langle m | \hat{F} | n \rangle - \frac{1}{2} 
  \frac{1}{\epsilon_n -\epsilon_m}  \langle m | [\hat{F},\hat{h}_{\rm ND}] | n \rangle .
\end{equation}
Again without any loss of generality we can take the diagonal part of the operator $\hat{G}$ to be zero. 

We now turn our attention to the third order correction. Using the Jacobi identity, after some algebra 
(\ref{11}) can be rewritten as 
\begin{eqnarray}
\label{19}
\left( - [\hat{G}_n, \hat{h}_{\rm D}] - [\hat{\Gamma}_n, \hat{H}_0 ] -  
\frac{1}{2} [\hat{H}_0, [\hat{F}_n, \hat{G}_n]] + \frac{1}{2} [\hat{F}^3_n, \hat{H}_0]  \right.  \nonumber \\
\left. + \frac{1}{2} (\hat{F}^2_n \hat{h}_{\rm D} + \hat{h}_{\rm D} \hat{F}^2_n ) 
- \hat{F}_n \hat{h} \hat{F}_n  \right) |n \rangle 
= \epsilon_n^{(3)}  |n \rangle .
\end{eqnarray}
Inspection reveals that the first four terms in (\ref{19}) lead to only non-diagonal operators. Hence the third order correction to the energy eigenvalues is given by
\begin{equation}
\label{20}
\epsilon_n^{(3)}  = \langle n | \frac{1}{2} (\hat{F}^2 \hat{h}_{\rm D} + \hat{h}_{\rm D} \hat{F}^2 )
- \hat{F}_n \hat{h} \hat{F}_n  |n \rangle , 
\end{equation}
which is again the standard result. The non-diagonal part of (\ref{19}) can then be used to explicitly calculate the operator $\hat{\Gamma}$. We omit that expression in this paper. We succeeded in showing that there exists an operator $\hat{S}$ that connects non-perturbed states to perturbed states in (\ref{4}) and that this operator is the same for all states. Even though there is extensive literature on the Dalgarno-Lewis method \cite{schw,mavro,nandi,Imbo:1983cd}, the existence of an operator that 
connects non-perturbed states to perturbed states, as well as the connection of this operator to the Dalgarno-Lewis method was not previously noted. 

Since the Dalgarno-Lewis method can be presented in terms of commutators, it is clearly very suitable for applications when Hamiltonians are formulated in an algebraic framework. We provide one such example in the next section. 

\section{Rotational Bands in Deep Potentials}

If the Hamiltonians considered have an underlying algebraic structure, then calculation of the operators 
$\hat{F}$, $\hat{G}$, etc. are considerably simplified. As an application we consider a three dimensional 
Hamiltonian with a deep potential, which is a function of $\mathbf{r}^2$ only. Such Hamiltonians arise in the study of alpha-clustering in nuclei. 

The low-lying states of the Hamiltonian
\begin{equation}
\label{21}
\hat{H} = \frac{\mathbf{p}^2}{2 \mu} + V (\mathbf{r}^2)
\end{equation}
can be evaluated by expanding the attractive potential in a power series in $\mathbf{r}^2$:
\begin{equation}
\label{22}
V (\mathbf{r}^2) = -V_0 ( 1 - \alpha \mathbf{r}^2 + \beta \mathbf{r}^4 + \cdots )
\end{equation}
Insight into the low-lying eigenstates of such a potential may be gained by taking the harmonic oscillator part 
\begin{equation}
\label{23}
\hat{H}_0 = \frac{\mathbf{p}^2}{2 \mu} + \alpha V_0 \mathbf{r}^2 
\end{equation}
as the unperturbed potential and 
\begin{equation}
\label{24}
\hat{h} = - V_0 \beta \mathbf{r}^4 
\end{equation}
as the perturbation \cite{Balantekin:1987ti}. Perturbation theory can be formulated algebraically introducing the harmonic oscillator creation and annihilation operators
\begin{equation}
\label{25}
\mathbf{a}^{\dagger}_i = \left( \frac{\mu \omega}{2 \hbar} \right)^{1/2} \mathbf{x}_i -  i \left( \frac{1}{2 \mu \hbar \omega} \right)^{1/2} \mathbf{p}_i  
\end{equation}
and
\begin{equation}
\label{25a}
\mathbf{a}_i = \left( \frac{\mu \omega}{2 \hbar} \right)^{1/2} \mathbf{x}_i + i \left( \frac{1}{2 \mu \hbar \omega} \right)^{1/2} \mathbf{p}_i  .
\end{equation}
To utilize and algebraic formalism one introduces the operators $K_+, K_-,  K_0$:
\begin{equation}
\label{26}
K_0 = \frac{1}{4} \sum_i \left( \mathbf{a}^{\dagger}_i \mathbf{a}_i + \mathbf{a}_i \mathbf{a}^{\dagger}_i \right). 
\end{equation}
\begin{equation}
\label{27}
K_+ =  \frac{1}{2} \sum_i \mathbf{a}^{\dagger}_i \mathbf{a}^{\dagger}_i , \>\>\>\>\>\> K_- =  (K_+)^{\dagger}, 
\end{equation}
which form the SU(1,1) Lie algebra with the commutation relations
\begin{equation}
\label{27a}
[K_0, K_{\pm}] = \pm K_{\pm},  \>\>\>\>\> [K_+, K_-] = -2K_0 .
\end{equation}
The Casimir operator of this algebra is
\begin{equation}
\label{28}
C_2 = K_0^2 - \frac{1}{2} \left( K_+K_- + K_- K_+ \right). 
\end{equation}
For the realization given in (\ref{26}) and (\ref{27}), this Casimir operator depends on the orbital angular momenta:
\begin{equation}
\label{29}
C_2 = \frac{1}{4} \mathbf{L}^2 - \frac{3}{16}. 
\end{equation}
The states are labeled by two numbers, $k$ and $m$.  
The eigenvalues of the Casimir operator are
\begin{equation}
\label{30}
C_2 | k,m \rangle = k (k-1) | k,m \rangle ,
\end{equation}
which implies that
\begin{equation}
\label{31}
k = \frac{1}{2} \left( \ell + \frac{3}{2} \right), 
\end{equation}
where $\ell$ is the orbital angular momentum quantum number. The label $m$ is the eigenvalue of the operator $K_0$:
\begin{equation}
\label{32}
K_0 | k,m \rangle = m |k,m \rangle.
\end{equation}
In the above equations $m$ takes values of $k, k+1, k+2, k+3, \cdots$. One also has
\begin{equation}
\label{33}
K_{\pm} | k,m \rangle = \left[ (m \pm k)( m \mp k \pm 1) \right]^{1/2} |k, m \pm 1 \rangle .
\end{equation}

It is easy to show that $\mathbf{r}^2$ can be expressed as
\begin{equation}
\label{34}
\mathbf{r}^2 = \frac{\hbar}{\mu \omega} ( K_+ + K_- + 2 K_0). 
\end{equation}
Hence the perturbation Hamiltonian of (\ref{24}) is
\begin{eqnarray}
\label{35}
\hat{h} = - \frac{\hbar^2 \beta}{2 \mu \alpha} ( K_+ + K_- + 2 K_0)^2 = 
- \frac{\hbar^2 \beta}{2 \mu \alpha} [ 6 K_0^2 - 2 C_2 + 2 (K_+ K_0 + K_0 K_+)  \nonumber \\
+2  (K_- K_0 + K_0 K_-) + K_+^2 + K_-^2 ] .
\end{eqnarray}
The first two terms in the above equation are diagonal in the unperturbed basis. It is then straightforward to verify that 
\begin{equation}
\label{36}
\hat{F} = - \frac{\hbar \beta}{2  \alpha \sqrt{2 \alpha \mu V_0}} 
\left[ 2 (K_- K_0 + K_0 K_-) -  2 (K_+ K_0 + K_0 K_+) + \frac{1}{2} 
(K_-^2 - K_+^2) \right] .
\end{equation}
To calculate the second order correction to the energy eigenvalues, (\ref{17}),  we need to calculate 
the diagonal part of the commutator of $\hat{F}$ with $\hat{h}_{ND}$. 
Defining the operators
\begin{equation}
\label{38}
\hat{A}_+ = K_+ K_0 + K_0 K_+ ,
\end{equation}
and
\begin{equation}
\label{39}
\hat{A}_- = K_- K_0 + K_0 K_- = (\hat{A}_+)^{\dagger} ,
\end{equation}
one sees, upon inspection, that there are two only diagonal contributions to $[\hat{F},\hat{h}_{ND}]$. 
One comes from the commutator
\begin{equation}
\label{40}
[\hat{A}_+, \hat{A}_-] = -16 K_0^3 + 8 C_2 K_0 - 2 K_0 ,
\end{equation}
and the other one from the commutator
\begin{equation}
\label{40b}
[K_+^2, K_-^2] = -2 (4 K_0^3 - 4 C_2 K_0 + 2 K_0).  
\end{equation}
Putting it all together we can write down the energy eigenvalues of the Hamiltonian of (\ref{21}) as 
\begin{eqnarray}
\frac{E_m}{V_0} &=& -1 + \left( m+\frac{1}{2} \right) \sqrt{2} \delta + \left( C_2 - 3 m^2 \right) \frac{\beta}{\alpha^2} \delta^2 
\nonumber \\ && 
- \left( 7 m C_2 - 15 m^3 -  \frac{3}{2} m \right) \sqrt{2} \frac{\beta^2 V_0}{\alpha^4} \delta^3 + \cdots 
\end{eqnarray}
where we defined the dimensionless expansion parameter
\begin{equation}
\delta = \hbar \sqrt{\frac{\alpha}{\mu V_0}}. 
\end{equation}
This Hamiltonian exhibits rotational bands (terms proportional to $C_2 \sim {\mathbf L}^2$) in 
its energy spectrum. The existence of such rotational bands for deep Gaussian potentials was proposed in \cite{buck} based on numerical calculations. We showed that application of the Dalgarno-Lewis method not only provides a convenient framework for calculating higher-order perturbative corrections, but also makes the existence of rotational bands manifest.

\ack
This work was supported in part
by the U.S. National Science Foundation Grant No. PHY-0855082 and  
in part by the University of Wisconsin Research Committee with funds
granted by the Wisconsin Alumni Research Foundation.


\vskip 1cm

\begin{thebibliography}{99}

\bibitem{DL}
Dalgarno A and Lewis J T 1956 \PRS {\bf A233} 70

\bibitem{schw}
Schwartz C 1959 \APNY {\bf 2} 156

\bibitem{mavro}
Mavromatis H A 1991 {\it Am.\ J. Phys.} {\bf 59}, 738

\bibitem{nandi}
Nandi T K, Bera P K, Panja M M, and Talukdar B 1996 \JPA {\bf 29}, 1101 

\bibitem{Imbo:1983cd}
  Imbo T and Sukhatme U 1984 
  {\it Am.\ J.\ Phys.}  {\bf 52} 140 

\bibitem{Balantekin:1987ti}
  Balantekin A B and Friedman W A 1987
  \PR {\bf C36 }  311 
  
\bibitem{buck}
Baldock R A, Buck B and Rubio J A 1984 {\it Nucl. Phys.} {\bf A426} 222; 
Buck B and Rubio J A, {\it J. Phys. G} 1984 {\bf 10} L209 (1984).
  

\end{thebibliography}
\end{document}